\documentclass{article}

\usepackage{arxiv}
\usepackage[utf8]{inputenc}  
\usepackage[T1]{fontenc}     
\usepackage{hyperref}        
\usepackage{url}             
\usepackage{booktabs}        
\usepackage{amsfonts}        
\usepackage{nicefrac}        
\usepackage{microtype}       
\usepackage{lipsum}
\usepackage{amsmath}
\usepackage{graphicx}
\usepackage{chngcntr}
\usepackage{cite}
\usepackage{comment}
\usepackage{xcolor}
\counterwithin{table}{section}

\title{Parkland Trauma Index of Mortality (PTIM): Real-time Predictive Model for PolyTrauma Patients}

\author{
Adam J. Starr, MD; Manjula Julka, MD; Arun Nethi; John D. Watkins MS; Ryan W. Fairchild, MD; \AND 
Michael W. Cripps, MD; Dustin Rinehart, MD; Hayden N. Box, MD
}
\date{October 2020}

\begin{document}
\maketitle

\keywords{Trauma mortality prediction\and polytrauma mortality prediction \and trauma predictive model \and machine learning model  }

\section*{Abstract}

\subsection*{Background} 
Vital signs and laboratory values are routinely used to guide clinical decision-making for polytrauma patients, such as the decision to use damage control techniques versus early definitive fracture fixation. Prior multivariate models have tried to predict mortality risk, but due to several limitations like one-time prediction at the time of admission, they have not proven clinically useful. There is a need for a dynamic model that captures evolving physiologic changes during patient's hospital course to trauma and resuscitation for mortality prediction.

\subsection*{Methods} 
The Parkland Trauma Index of Mortality (PTIM) is a machine learning algorithm that uses electronic medical record (EMR) data to predict $48-$hour mortality during the first $72$ hours of hospitalization. The model updates every hour, evolving with the patient's physiologic response to trauma. The model was trained on $1935$ hospitalized polytrauma patient encounters from $2009 - 2014$ and was tested on $516$ encounters from $2015 - 2016$. Area under (AUC) the receiver-operator characteristic curve (ROC), sensitivity, specificity, positive (PPV) and negative predictive value (NPV), and positive and negative likelihood ratios (LR) were used to evaluate model performance. 

\subsection*{Results} 
The PTIM accurately predicted $52$ of the $63$ $12-$hour time intervals within $48$ hours of mortality, for a sensitivity of $82.5\%$ ($95\%$ CI $[73.1\%, 91.9\%]$). The specificity was $93.6\%$ ($95\%$ CI $[92.5\%, 94.8\%]$), and the PPV was $32.5\%$ ($95\%$ CI $[25.2\%, 39.7\%]$). The model predicted survival for $1608$ time intervals and was incorrect 11 times, yielding a NPV of $99.3\%$ ($95\%$ CI $[98.9\%, 99.7\%]$). The positive LR was $12.9$ ($95\%$ CI $[9.9, 14.6]$) and the negative LR was $0.19$ ($95\%$ CI $[0.10, 0.31]$). The AUC of the ROC curve was $0.94$. Model performance was stable over the first $72$ hours of hospitalization. 

\subsection*{Conclusion} 
By evolving with the patient's physiologic response to trauma and relying only on EMR data, the PTIM overcomes many of the limitations of prior mortality risk models. It may be a useful tool to inform clinical decision-making for polytrauma patients early in their hospitalization.

\section{Introduction}

\subsection{Background}
The timing of fracture fixation in polytraumatized patients has been an evolving concept over the past several decades \cite{d2013evolution}. In the $1980$s, it was demonstrated that polytrauma patients had poorer outcomes and higher incidence of respiratory complications with delayed fixation \cite{bone1989early, johnson1985incidence, lozman1986pulmonary}. However, a subset of patients benefits from a less aggressive approach \cite{jaicks1997early,pape1993primary,townsend1998timing}. These ``borderline" patients cannot tolerate the added physiologic insult and can potentially develop an exaggerated inflammatory response, acute respiratory distress syndrome, and multi-organ failure in response to prolonged surgery \cite{pape1993primary,faist1983multiple, harwood2005alterations, pape2000biochemical, waydhas1996posttraumatic}. Damage Control Orthopedics (DCO) emerged as the accepted initial approach to fractures in polytrauma patients \cite{scalea2000external}, and by the early $2000$s, multiple studies demonstrating the benefits of damage control shifted the pendulum from early total care towards temporizing measures in severely injured patients \cite{pape1999optimal,pape2003impact, pape2002changes, taeger2005damage}. The concept of Early Appropriate Care (EAC) emerged in the late $2000$s, representing a compromise between damage control and ETC that minimizes the extent of the initial surgery while allowing mobilization and continued resuscitation \cite{nahm2011early, pape2007impact}.  

While differentiating stable patients and those in extremis is generally not problematic, it is difficult in the acute clinical setting to reliably and accurately assign mortality risk to those patients who fall into the borderline category. Conventional methods of determining patient stability and adequacy of resuscitation rely on vital signs, such as heart rate, respiratory rate, and systolic blood pressure, as well as laboratory values, such as base deficit, lactate. Various multivariable scoring systems, such as the injury severity score (ISS), revised trauma score (RTS), trauma and injury severity score (TRISS) have attempted to systematize the classification of anatomic injury severity and the degree of physiologic derangement in order to predict mortality risk. These scoring systems, though useful in retrospective review, are cumbersome and impractical in the acute clinical setting. Multiple efforts to systematically identify at-risk patients and predict outcomes have been made \cite{hildebrand2015development, kunitake2018trauma}. However, models based on a single time point do not allow for changes in resuscitation status. Furthermore, the primary end point, in-hospital mortality, is predicted by a static score at the time of admission that does not evolve with the patient's hospital course. This results in a vague prediction of the patient's outcome that is difficult to apply in the clinical setting.   

We have found no clinically useful mortality risk model that provides a well-defined prediction interval and updates with the patient's evolving physiologic response to trauma and resuscitation. The purpose of our study was to use electronic medical record (EMR) data and machine learning techniques to develop an automated algorithm that predicts $48-$hour mortality in polytraumatized patients over the first $72$ hours of their hospitalization.

\section{Methods}

\begin{flushleft}\emph{Data}\end{flushleft}

\begin{figure}[h!]
  \centering
  \includegraphics[width = .6\textwidth]{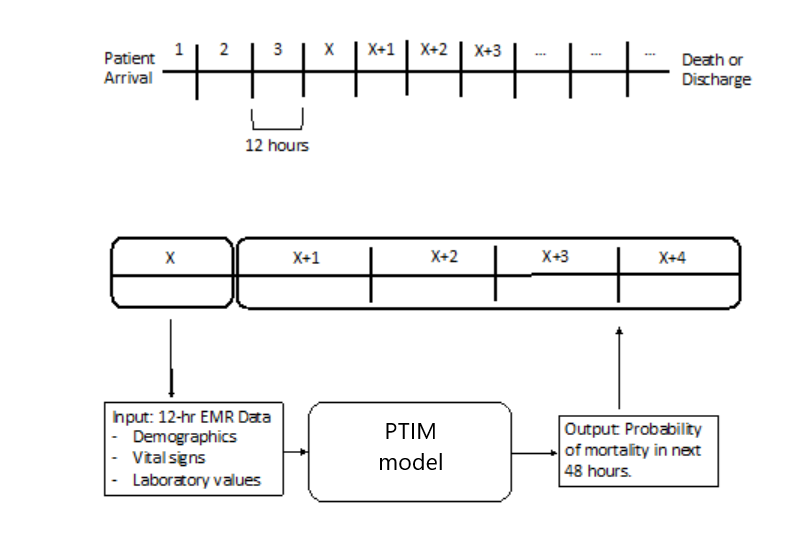} 
  \caption{PTIM schematic. (Top) Patient encounters were divided into time intervals from presentation to death or discharge. (Bottom) For each time interval, electronic medical record data was used as PTIM input to predict morality in the next 48 hours using preceding 12-hour data.}
  \label{fig:PTIM_schematic}
\end{figure}

After Institutional Review Board approval, the Electronic Medical Record (EMR) data at a large, urban level I trauma center was queried for the years $2009 - 2016$. There were $2451$ unique patient encounters during this time period with the inclusion criteria of presentation from $2009$ to $2016$, age of at least 18 years, time since arrival at least 12 hours and an exclusion of any burn patients. For each hourly output, our machine learning algorithm used preceding $12-$hour EMR data to predict the probability of mortality in the next $48$ hours. Each patient encounter was divided into $12-$hour time intervals, from presentation to time of discharge or death (Figure \ref{fig:PTIM_schematic}). The PTIM training set consisted of $1935$ unique patient encounters from $2009$ to $2014$ broken down into $10603$ time intervals. The test set was comprised of 516 unique patient encounters from $2015-2016$ with $1768$ time intervals. The data was imbalanced with survival-mortality ratio of $46$:$1$.

\begin{flushleft}\emph{Algorithm } \end{flushleft}

Bagging (known as Bootstrap Aggregating) is an ensemble technique used for classification or regression and was proposed by Leo Breiman \cite{breiman2001random}. The machine learning meta-estimator fits base estimators (e.g. decision trees) on random subsets of the original data and then uses aggregation (voting or averaging) to generate a prediction. Bagging technique is a way to decrease the variance and improve the model stability. 
 
\begin{flushleft}\emph{PTIM model algorithm} \end{flushleft}
 
\begin{figure}[h!]
  \centering
  \includegraphics[width = .6\textwidth]{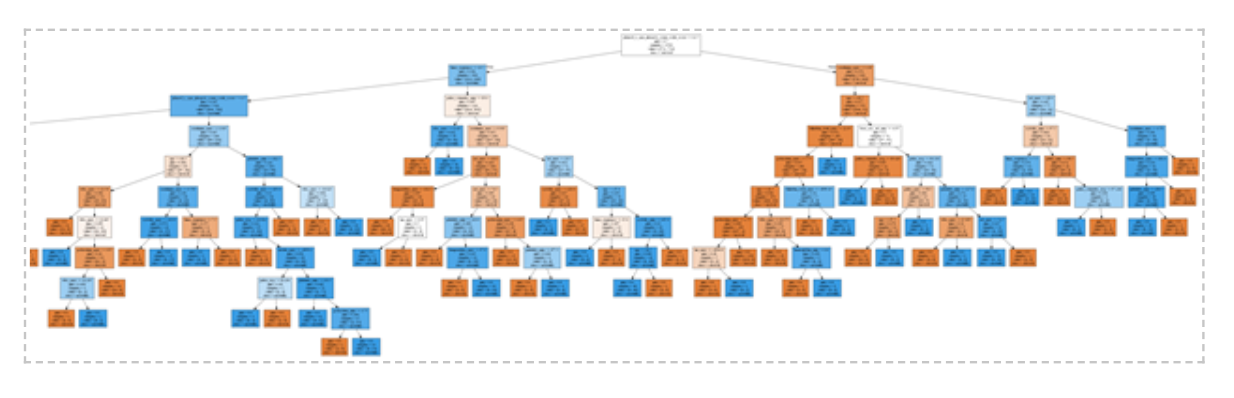} 
  \caption{An example of one of the many decision trees that makes up the PTIM.}
  \label{fig:example_PTIM}
\end{figure}

The Parkland Trauma Index of Mortality (PTIM) is a balanced bagging ensemble of decision tree classifiers. An example of one of the decision trees included in the model is shown in the Figure \ref{fig:example_PTIM}. The imbalance in data set is handled by the random under sampling technique of balanced bagging classifier.

\begin{flushleft}\emph{PTIM feature set } \end{flushleft}

The PTIM is comprised of the features shown in the Table \ref{tab:features}. These were chosen by a fellowship-trained attending orthopedic trauma surgeon and a fellowship-trained attending critical care trauma surgeon. The features were chosen by their clinical utility and availability in the EPIC (Verona, WI) electronic medical record (EMR) that has been in use at our institution since $2009$.  

While the feature Glasgow Coma Score (GCS) is based on the most recent value recorded, all other labs and vitals used as features are aggregated using the preceding $12-$hour data collected from the EMR for each hourly prediction. Time since arrival is the categorical variable of the time from admission. Missing values were handled with feed forward filling for $24$ hours. For greater than $24$ hours, arbitrary value imputation was used. 

\begin{table}[h!]
\caption{Features included in the Parkland Trauma Index of Mortality.}  
\label{tab:features}
\begin{center}
\begin{tabular}{| p {2cm}| p {3cm}| p {2cm}| p {2cm}| p{1.5cm} |} 
 \hline 
 Vital Signs & Laboratory Values & Clinical Scores & Demographics & Time \\
 \hline 
Heart rate & Base deficit & Glasgow coma score & Age &Time since arrival  \\ [0.5ex] 
 \hline
 Systolic blood pressure & Lactate & & & \\
 \hline
 Blood oxygen saturation & Hemoglobin concentration & & & \\ 
\hline
Body temperature   & Platelet count & & & \\ 
\hline
& White Blood Cell count & & & \\
\hline
& International normalized ratio  & & & \\
\hline
& Potassium & & &\\
\hline
& Aspartate aminotransferase (AST)  & & & \\
\hline
& Total Bilirubin & & &\\
\hline
& Albumin  & & &\\
\hline
& Creatinine  & & &\\
\hline
\end{tabular}
\end{center}
\end{table}

\begin{flushleft}\emph{Training } \end{flushleft}
 
The training and test sets were categorized based on year of presentation. The training set consisted of patient encounters from $2009$ to $2014$.The model was trained using group k-fold cross-validation to have all the records related to a patient in one single fold and to avoid any chances of data leakage. The model was then evaluated on the test set of patient encounters from $2015$ to $2016$. 

\begin{flushleft}\emph{Evaluation } \end{flushleft}

Area under (AUC) the receiver operator characteristic (ROC) curve, sensitivity, specificity, positive and negative predictive values, as well as positive and negative likelihood ratios with $95\%$ confidence intervals were calculated.

\section{Results}

\begin{figure}[h!]
  \centering
  \includegraphics[width = .8\textwidth]{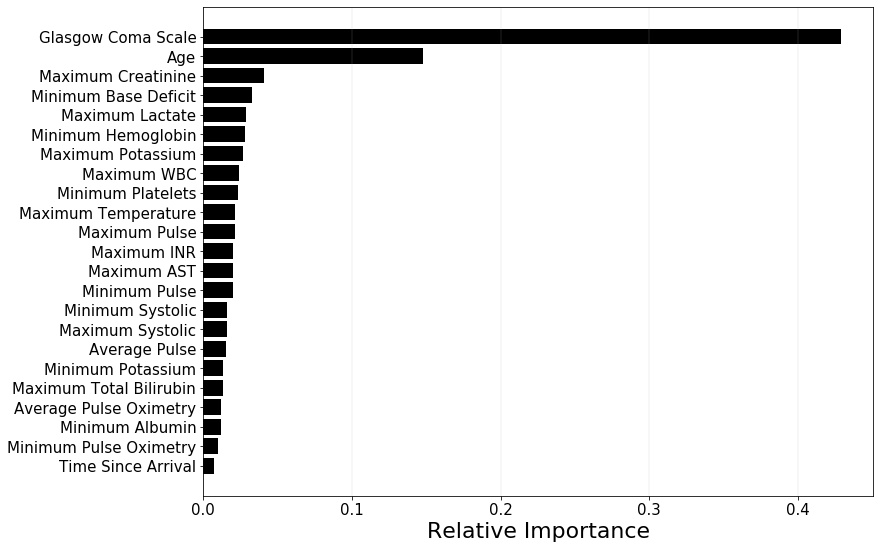} 
  \caption{PTIM relative feature importance.}
  \label{fig:FI}
\end{figure}

Relative feature importance of the model is shown in the Figure \ref{fig:FI}. Minimum Glasgow coma score (GCS) had the greatest influence on our model with a relative feature importance of $42.9\%$ and patient's Age was the next highest influence with a relative feature importance of $14.8\%$. Several model features had relative importance between $1-5\%$. These included base deficit, white blood cell count, international normalized ratio (INR), minimum systolic blood pressure, average pulse, average oxygen saturation, hemoglobin concentration, maximum body temperature, lactate concentration, maximum pulse, maximum systolic blood pressure, maximum serum potassium concentration, minimum pulse, and maximum aspartate aminotransferase (AST). Features with relative importance of less than $1\%$ included time since admission.  

\begin{figure}[h!]
	\centering
    \includegraphics[width = .7\textwidth]{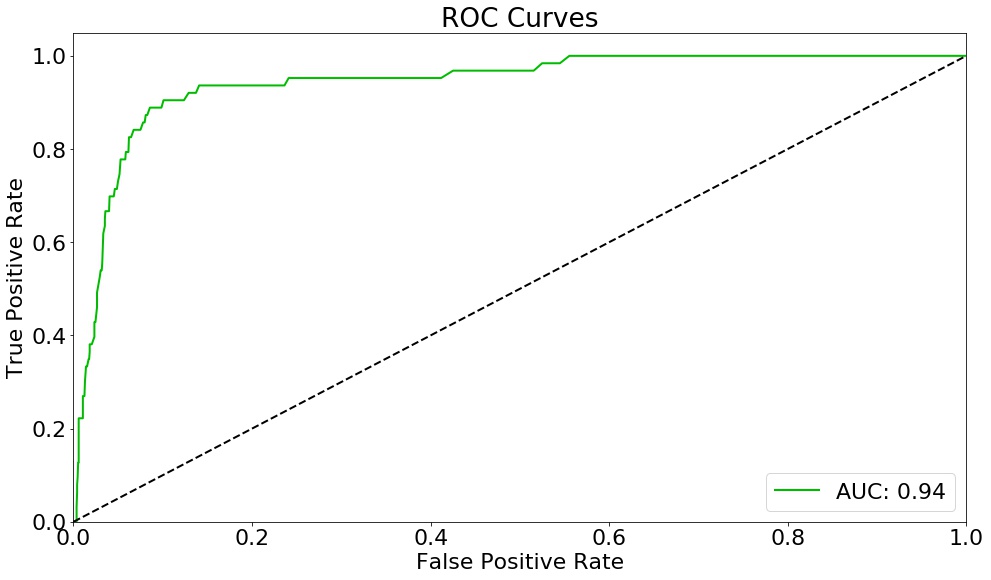}
	\caption{ROC Curve}
	\label{fig:ROC_PR1}
\end{figure}

\begin{figure}[h!]
	\centering
    \includegraphics[width = .7\textwidth]{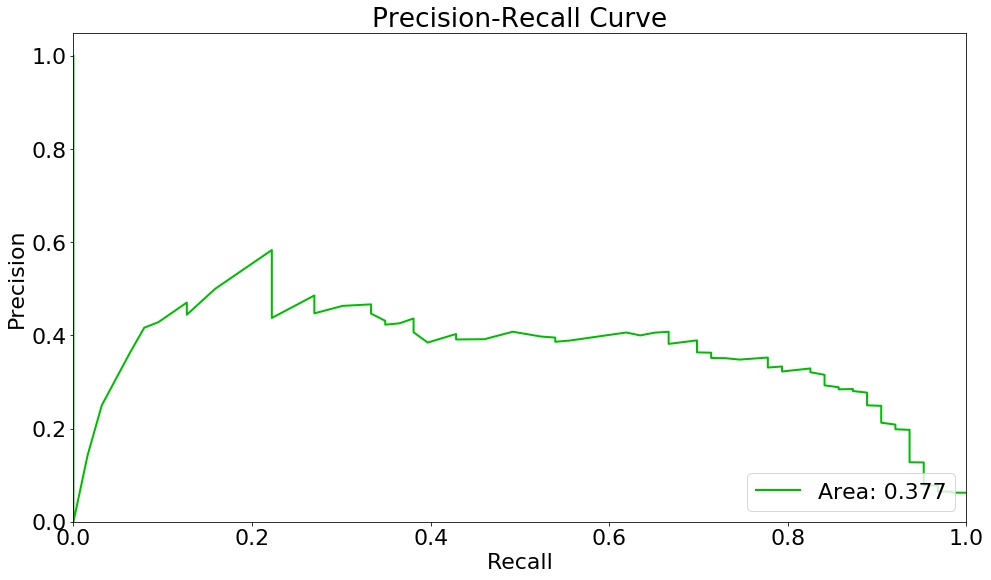}
	\caption{Precision Recall Curve.}
	\label{fig:ROC_PR2}
\end{figure}

Raw model performance and performance measures are shown in the Tables \ref{tab:performance} and \ref{tab:performance_measurement}. The PTIM accurately predicted $52$ of the $63$ time intervals within $48$ hours of mortality, yielding a sensitivity of $82.5\%$ ($95\%$ CI $[73.1\%, 91.9\%]$). The specificity was $92.2\%$ ($95\%$ CI $[90.8\%, 93.6\%]$). The PTIM predicted $48-$hour mortality in $160$ time intervals and was correct for $52$ of them, for a positive predictive value of $32.5\%$ ($95\%$ CI $[27.7\%, 36.0\%]$). The model predicted 48-hour survival for 1608 time increments and was wrong 11 times, for a negative predictive value of $99.7\%$ ($95\%$ CI $[99.2\%, 99.9\%]$). The positive likelihood ratio was $12.9$ ($95\%$ CI $[9.9 - 14.6]$). The negative likelihood ratio was $0.19$ ($95\%$ CI $[0.11, 0.31]$). These performance measures were stable over the first $72$ hours of hospitalization. The area under the curve (AUC) for the receiver operating characteristic curve was $0.94$ (Figure \ref{fig:ROC_PR1}, \ref{fig:ROC_PR2}).  

\begin{table}[h!]
\caption{PTIM predictions vs Actuals}  
\label{tab:performance}
\begin{center}
\begin{tabular}{|c | c | c|} 
 \hline 
 \multicolumn{3}{|c|}{Actual} \\
 \hline
Predicted & Death & Survival \\ [0.5ex] 
 \hline
 Death & $52$ & $108$ \\ 
 \hline
 Survival & $11$ & $1597$ \\ 
 \hline
\end{tabular}
\end{center}
\end{table}

\begin{table}[h!]
\caption{PTIM performance measurements.}  
\label{tab:performance_measurement}
\begin{center}
\begin{tabular}{|c | c | c|} 
\hline 
 & Value & 95\% CI \\[2ex] 
\hline
Sensitivity & 82.5\% & 73.1-91.9\%  \\ [2ex] 
\hline
Specificity & 93.6\% & 92.5-94.8\% \\[2ex] 
\hline
PPV & 32.5\% & 25.2-39.7\% \\[2ex] 
\hline
NPV  & 99.3\% & 99.2-99.9\% \\[2ex] 
\hline
LR+ & 12.9 & 9.9 - 14.6 \\[2ex] 
\hline
LR- & 0.19 & 0.10 – 0.31 \\[2ex] 
\hline
\end{tabular}
\end{center}
\end{table}

\section{Discussion}

The PTIM is a machine learning algorithm that uses EMR data to predict $48-$hour mortality in polytraumatized patients during the first $72$ hours of their hospitalization. With a positive likelihood ratio of $>10$ and negative likelihood ratio of $<0.2$, the PTIM may be useful when attempting to assess a patient's risk of $48-$hour mortality. 

Prior authors have described other mortality risk scoring models. Hildebrand et al published a scoring system called the PolyTrauma Grading Score (PTGS) \cite{hildebrand2015development}. It uses the systolic blood pressure as well as laboratory values such as base deficit, INR, and platelet count. The model requires the calculation of the new injury severity score (NISS) and the number of packed red cells the patient has received. These data points are taken from when the patient first presents to the emergency department. From these parameters, patients are assigned a score, and from this score, the patient can be described as stable ($<5\%$ mortality risk), borderline ($15\%$ mortality), and unstable conditions ($40\%$ mortality).  

Kunikate et al described the Trauma Early Mortality Prediction Tool (TEMPT) in $2018$. \cite{kunitake2018trauma} They derived a model independent of the abbreviated injury severity (AIS) score to assess trauma mortality risk. Utilizing multiple logistic regression analysis, they identified presence of traumatic brain injury, age $> 59.5$ years, INR $> 1.25$, PTT $> 31.4$ s, base excess < $-4.35$ mmol/L and temperature $< 36.25$ C as independent predictors of mortality. The area under the curve of the receiver operator characteristic curve for predicting $28-$day mortality was $0.94$ ($95\%$ CI $[0.92, 0.97]$).  

Our model supports findings from prior studies that traumatic brain injury, age, coagulopathy (low platelet count and increased INR), and poor tissue perfusion (increased base deficit and lactate) are major predictors of mortality. However, our model overcomes the practical limitations of prior models. By only looking at a patient's vital signs and laboratory values upon arrival to the trauma bay, prior models such as the TEMPT and PTGS, do not consider a patient's evolving physiologic response to trauma and ongoing resuscitation.  Our model updates every hour, and in doing so, can evolve with the patient's physiologic status over the first $72$ hours of hospitalization. Another limitation of prior models is their cumbersome nature. The TEMPT and the PTGS require clinicians to compare parameters to defined thresholds and then calculate a score. In the case of the PTGS, the clinician has to calculate a NISS, which can be difficult to determine soon after trauma when the severity of injuries may be not be known. Our model was built as an automated, machine learning algorithm that requires no input from the clinician. Moreover, our model is independent of anatomic injury scores, such as the ISS and NISS, relying only on data present in the EMR. 

Prior models have described the risk of ``in-hospital" mortality. This poorly defined time period may refer to death in the first $24$ hours or first $30$ days after injury. Our mortality model scans the EMR for the preceding $12$ hours and gives a mortality prediction for the next $48$ hours. We thought this provided clinicians with a more clearly defined description of mortality risk. Another strength of our model is its discriminative ability. It can sort patients into low-risk and high-risk groups. Based on the negative predictive value of our model, if our model predicted survival, then the risk of mortality was $0.7\%$. However, if the model predicted mortality, then the risk of $48-$hour mortality was $32\%$, almost $1$ in $3$.  

Our study is not without limitations. First, this is a single institution study. We relied on objective laboratory and vital sign parameters that are accessible from the EMR in order to limit the institutional bias, but different clinical practices from institution to institution may alter the performance of the model. Second, our machine learning algorithm is large assemblage of decision trees, and as such is less intuitive than commonly used multivariate models, such logistic regression models. Our rationale for utilizing a less intuitive machine learning algorithm was that these more sophisticated algorithms are better suited to discern more complex relationships between physiologic variables. Third, this is a retrospective study, and as such does not evaluate the model's real-time performance.  We tried to mimic real-time incorporation by training in the model on $2009-2014$ data and testing its performance on data from $2015-2016$. The model's real-time implementation has been via a secure cloud platform Isthmus \cite{arora2019isthmus} and has been running live for the clinical decisions at level $1$ trauma center hospital since August $2019$. The Parkland Trauma Index of Mortality model is a free software and is distributed under the terms of the GNU Lesser General Public License (LGPL). 

The Parkland Trauma Index of Mortality (PTIM) is an automated machine learning algorithm that predicts $48-$hour mortality in polytrauma patients. With a positive likelihood ratio of greater than $10$ and negative likelihood ratio of less than $0.1$, the PTIM may be a useful tool for critical care providers and orthopaedic surgeons to guide clinical decision-making early in the hospitalization.

\section{Acknowledgment}
Parkland Health and Hospital System (PHHS); PHHS IT team; Parkland Center for Clinical Innovation(PCCI); PCCI Data Science and Technology team

\bibliographystyle{abbrv}
\bibliography{reference}

\end{document}